\documentclass[preprint,showpacs,preprintnumbers,amsmath,amssymb,superscriptaddress]{revtex4}
\usepackage{graphicx}
\usepackage{dcolumn}
\usepackage{bm}
\usepackage{hyperref}
\usepackage{multirow}
\usepackage{amsmath}

\begin{document}

\title{Search for decays of the $^{9}$B nucleus and Hoyle state in $^{14}$N nucleus dissociation}

\author{E. Mitsova}
\affiliation {Joint Institute for Nuclear Research, Dubna, Russia} 
\author{A.A. Zaitsev}
\email{zaicev@lhe.jinr.ru}
\affiliation {Joint Institute for Nuclear Research, Dubna, Russia} 
\author{D.A. Aretemenkov}
\affiliation {Joint Institute for Nuclear Research, Dubna, Russia} 
\author{N.K. Kornegrutsa}
\affiliation {Joint Institute for Nuclear Research, Dubna, Russia} 
\author{V.V. Rusakova}
\affiliation {Joint Institute for Nuclear Research, Dubna, Russia} 
\author{R. Stanoeva}
\affiliation {Southwestern University, Blagoevgrad, Bulgaria}
\author{P.I. Zarubin}
\affiliation {Joint Institute for Nuclear Research, Dubna, Russia} 
\author{I.G. Zarubina}
\affiliation {Joint Institute for Nuclear Research, Dubna, Russia}

\begin{abstract}
First results of an analysis to determine contribution of decays of the unstable $^{8}$Be and $^{9}$B nuclei and the Hoyle 3$\alpha$-state to dissociation of $^{14}$N $\to$ 3He (+H) are presented. As the research material, layers of nuclear track emulsion longitudinally exposed to 2.9 $A$ GeV/$c$ $^{14}$N nuclei with at the JINR Nuclotron. Under the assumption that the He and H fragments retain momentum per nucleon of the primary nucleus, these unstable states are identified by the invariant mass calculated from the emission angles of the fragments.
\end{abstract}


\maketitle

\section{Introduction}
\label{intro}
In the region of limiting fragmentation a primary energy value or target composition do not affect on manifestation of cluster features of nuclei under study. The study of the composition of relativistic dissociation of light nuclei by the nuclear track emulsion method (NTE) in the BECQUEREL experiment, indicates a contribution of the unstable nucleus decays $^{8}$Be $\to$ 2$\alpha$ and $^{9}$B $\to$ $^{8}$Be$p$, as well as the 3$\alpha$ decay of the Hoyle state (HS) \cite{1,2}. Following its identification in the dissociation of the $^{12}$C nucleus, HS manifested in the $^{16}$O case as an unstable 3$\alpha$ nucleus similar to $^{8}$Be one. According to the widths of $^{8}$Be (5.6 eV), $^{9}$B (540 eV), and HS (9.3 eV) \cite{3}, their decays occur on times which are few orders of magnitude greater than relativistic reaction ones. The minimum decay energy of $^{8}$Be (92 keV), $^{9}$B (185 keV), and HS (378 keV) \cite{3} is shown in the smallest opening angles of a pair and triples of He and H fragments distinguishing them from other fragmentation products.

The invariant mass of a system of relativistic fragments is defined as the sum of all products of 4-momenta $P_{i,k}$ of fragments $M^{*2}$ = $\Sigma$($P_i\cdot P_k$). Subtracting the mass of the sum of the masses of the fragments $Q$ = $M^*$ - $M$ is convenient for presentation. The components $P_{i,k}$ are determined in the approximation of conservation of the initial momentum per nucleon by fragments. Often it is sufficient to assume that He - $^{4}$He and H - $^{1}$H correspond, since in the case of stable nuclei the established contributions of $^{3}$He and $^{2}$H do not exceed 10\%. This simplification, used below, is especially true in the case of extremely narrow $^{8}$Be and $^{9}$B decays. Under the assumption that the fragments of momentum per nucleon of the primary nucleus are retained, the unstable states are identified by the minimum invariant mass calculated from the angles of emission of He and H fragments. Then, for their selection, it is sufficient to introduce a limitation of several hundred keV on the invariant mass of the final pairs and triples (minus the total mass of the latter).

Being independent of the initial energy and the multiplicity of dissociation the developed approach makes it possible to study uniformly the role of the unstable nuclei as elements of the cluster structure of light nuclei. Moreover, there is a prospect of searching for nuclear-molecular states with electromagnetic widths such as $^{9}$B$p$, $^{9}$B$\alpha$ and HS$p$, the products of $\alpha$-particle or proton decay of which could serve as HS or $^{9}$B, and ultimately $^{8}$Be. This gives rise to the prospect of ``unraveling in the reverse order'' of cascade decays against the background of other products of nuclear dissociation.

The material for such a study can be NTE layers longitudinally exposed to $^{14}$N 2.9 $A$ GeV/$c$ nuclei at the JINR Nuclotron in 2004. The charge topology of $^{14}$N dissociation is established earlier, and the $^{8}$Be contribution is estimated at 25-30\% \cite{4}. The role of $^{9}$B and HS remains a topical issue. The $^{14}$N nucleus, following $^{10}$B, belongs to an extremely small number of stable odd-odd nuclei (also $^{2}$H, $^{6}$Li, $^{50}$V). The similarity of the cluster pattern of dissociation of these nuclei in general and formation of the unstable nuclei in particular is possible. The data on the dissociation of the $^{10}$B nucleus obtained on the basis of exposure at the Nuclotron in 2002 provide the closest possibility of verification and comparison \cite{5}. The progress of analysis in this direction is presented below.

\begin{figure}
	\resizebox{1\textwidth}{!}{%
		\includegraphics{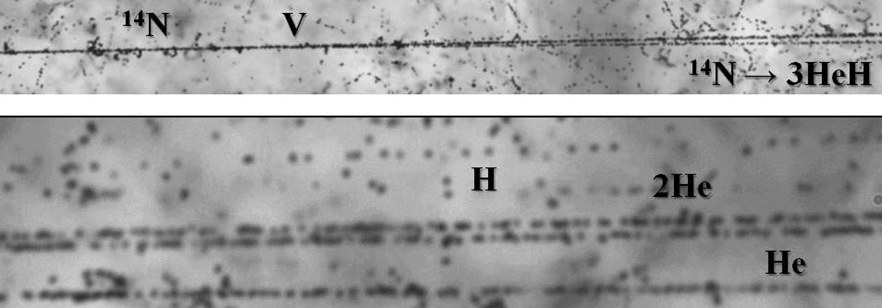}
	}
	\caption{Closeup of coherent dissociation of 2.9 $A$ GeV/$c$ $^{14}$N nucleus. Grain size no more than 0.5 microns. Top photo on the left shows $^{14}$N track accompanied by short tracks of $\delta$-electrons; position of vertex with a sharp decrease in ionization is marked (V). At 1 mm in fragment direction their tracks are distinguishable (bottom photo); narrow 2He pair corresponds to $^{8}$Be decay.}
	\label{fig:1}       
\end{figure}

\section{INTERACTION SEARCH AND MEASUREMENTS}
\label{sec:1}
The search for fragmentation events is carried out by transverse scanning of the NTE layers. The observed He track pairs or pairs with heavier fragments were scanned to an interaction vertex to find the remaining tracks. Table \ref{tab:1} shows the distribution the topology of fragments with a charge $Z$ for the main statistics. Although there is a difference from the results of the search for events along primary traces \cite{4}, the main branch of the $^{14}$N $\to$ 3He statistics is observed with the same efficiency. An example of the event of coherent dissociation $^{14}$N $\to$ 3HeH (``white'' star), not accompanied by produced mesons and target fragments in a wide cone, is shown in Fig. 1. The further analysis does not include a small number of ``white'' 3$\alpha$ stars assigned to an admixture of beam $^{12}$C nuclei.

\begin{table}
	\caption{Distribution of ``white'' stars $N_\text{ws}$ and peripheral interactions with target fragments $N_\text{tf}$ over the number of relativistic fragments $N_Z$ with charge $Z$.} 
	\label{tab:1}       
	\begin{tabular}{|c|c|c|c|c|c|c|}  \hline
		$\Sigma Z$ & \multicolumn{2}{c|}{6} & \multicolumn{4}{c|}{7} \\ \hline
		$N_{Z>2}$ & -- & -- & 1 & -- & -- & -- \\ \hline
		$N_{Z=2}$ & 3 & 2 & 2 & 2 & 3 & 3 \\ \hline
		$N_{Z=1}$ & -- & 2 & -- & 3 & 1 & -- \\ \hline
		$N_\text{ws}$ & 11 & 9 & 9 & 16 & 59 & -- \\ \hline
		$N_\text{tf}$ & 90 & 138 & 11 & 81 & 167 & 90 \\ \hline
	\end{tabular}
\end{table}

The statistics of the dissociation channel 3HeH which is of primary interest is leading both among the ``white'' stars $N_\text{ws}$(3HeH) and interactions with target fragments $N_\text{tf}$(3HeH). For events with 3HeH and 3He fragments, which look 
like a reaction of a neutron or proton stripping with destruction of the target, there is a significant difference in the numbers $N_\text{tf}$. Their ratio is $N_\text{tf}$(3HeH)/$N_\text{tf}$(3He)  = 1.9 $\pm$ 0.2. A similar channel of 2HeH dissociation of the isotope $^{10}$B, also leading, was represented by the number of stars of both types $N_\text{ws}$(2HeH) = 103 (76\% $N_\text{ws}$) and $N_\text{tf}$(2HeH) = (48\% $N_\text{tf}$), and the ratio is $N_\text{tf}$(2HeH)/$N_\text{tf}$(2He) = 182/89 or 2.1 $\pm$ 0.3 \cite{5}. The above ratios can be used to check detailed calculations of the structure of these nuclei.

With the aim of reconstructing decays of the unstable nuclei by the most precise coordinate method, measurements of the planar and depth angles of the tracks of relativistic fragments have been started using the described statistics. To date, measurements have been made in 60 3HeH events, including 14 ``white'' stars, and 20 3He events with tracks in a wide cone. On this basis, the polar angles of these tracks relative to the primary ones $\theta_\text{He}$ and $\theta_\text{H}$ are calculated. Their distributions are shown in Fig. 2 are described by the Rayleigh distribution with the parameter values $\sigma_{\theta_\text{He}}$ = 9.6 $\pm$ 0.5 mrad and $\sigma_{\theta_\text{H}}$ = 25 $\pm$ 9 mrad.

\begin{figure}
	\resizebox{1\textwidth}{!}{%
		\includegraphics{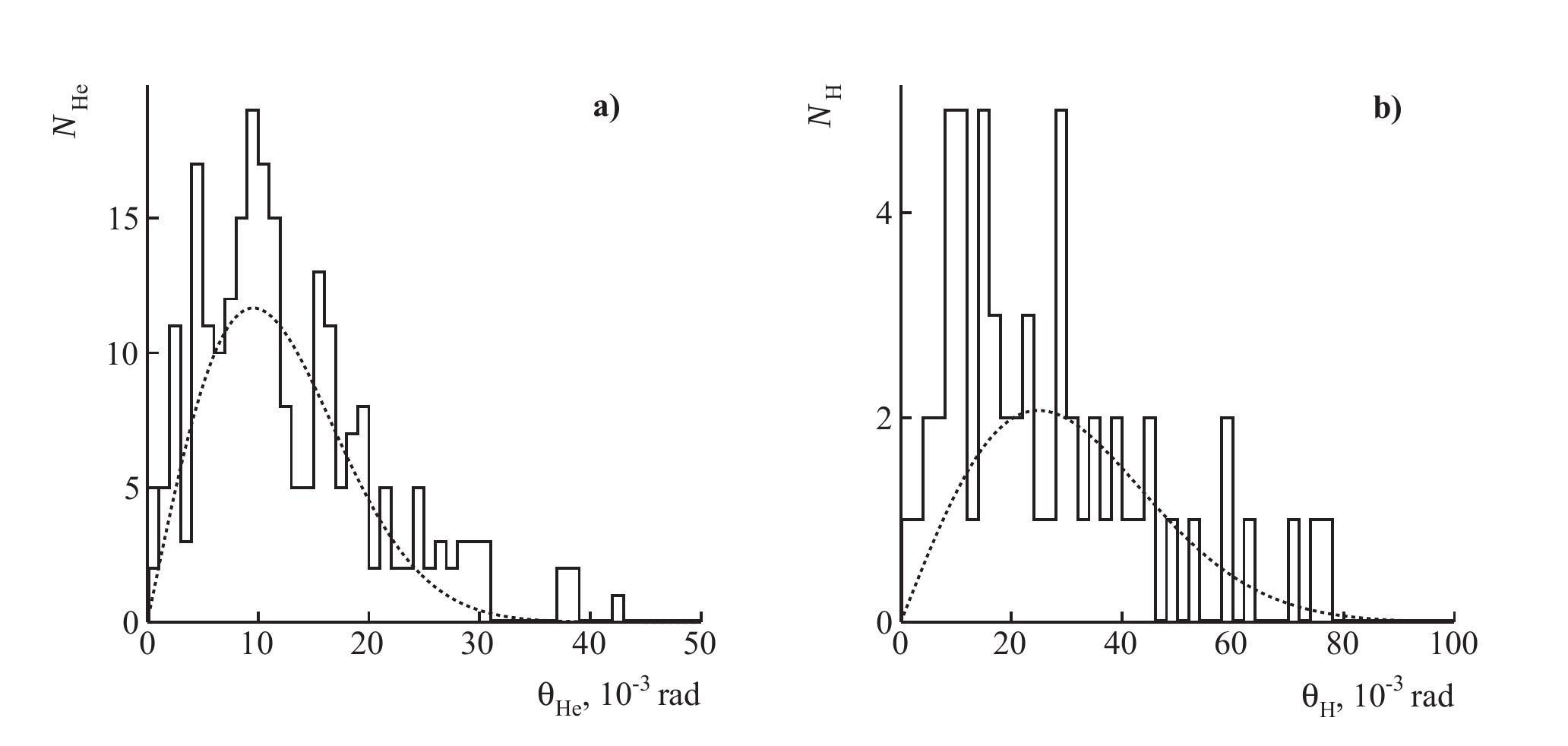}
	}
	\caption{Distribution of fragments from dissociation $^{14}$N $\to$ 3He + H over polar angles $\theta_\text{He}$ (a) and $\theta_\text{H}$ (b); points – Rayleigh distribution.}
	\label{fig:2}       
\end{figure}

\begin{figure}
	\resizebox{0.6\textwidth}{!}{%
		\includegraphics{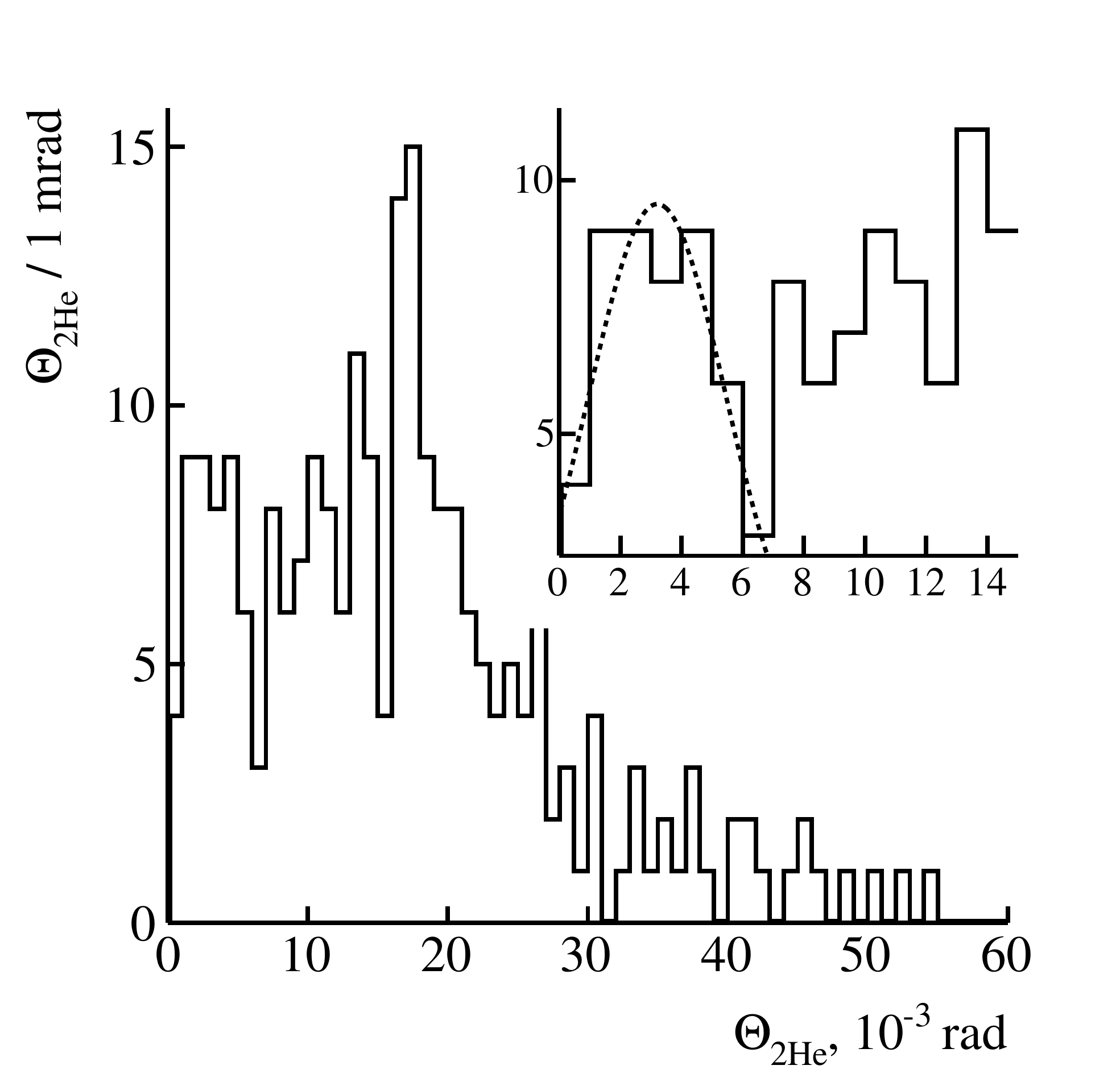}
	}
	\caption{Distribution over the relative spatial angle $\Theta_\text{2He}$ for combinations of pairs of $\alpha$-particles in events $^{14}$N $\to$ 3$\alpha$ (+ H); inset:  enlarged region of the lowest values of $\Theta_\text{2He}$; points  – Gaussian approximation.}
	\label{fig:3}       
\end{figure}

\section{IDENTIFICATION OF UNSTABLE STATES}
\label{sec:2}

The angle measurements allow one to go to the inter-track angle analysis. The distribution of combinations of pairs of $\alpha$-particles over the relative spatial angle $\Theta_\text{2He}$ in events $^{14}$N $\to$ 3He (including + H) is shown in Fig. 3. In the region $\Theta_\text{2He}$ $<$ 6 mrad, the average value is  $\left\langle \Theta_\text{2He} \right\rangle$  = 3.2 $\pm$ 0.4 mrad at RMS 2.2 mrad. According to the distribution, it corresponds to the decays of the $^{8}$Be ground state in the region of the lowest values of the invariant mass of $\alpha$-pairs $Q_{2\alpha}$ (Fig. 4). For 30 events, the average value of $\left\langle Q_\text{2He} \right\rangle$ satisfying the condition $Q_{2\alpha}$ ($^{8}$Be) $<$ 0.2 MeV was 77 $\pm$ 9 keV at RMS 56 keV. On this basis, the contribution of $^{8}$Be nuclei among the $^{14}$N $\to$ 3He events is 38 $\pm$ 7\% and 45 $\pm$ 9\% in the $^{14}$N $\to$ 3$\alpha$ + H channel.

\begin{figure}
	\resizebox{0.6\textwidth}{!}{%
		\includegraphics{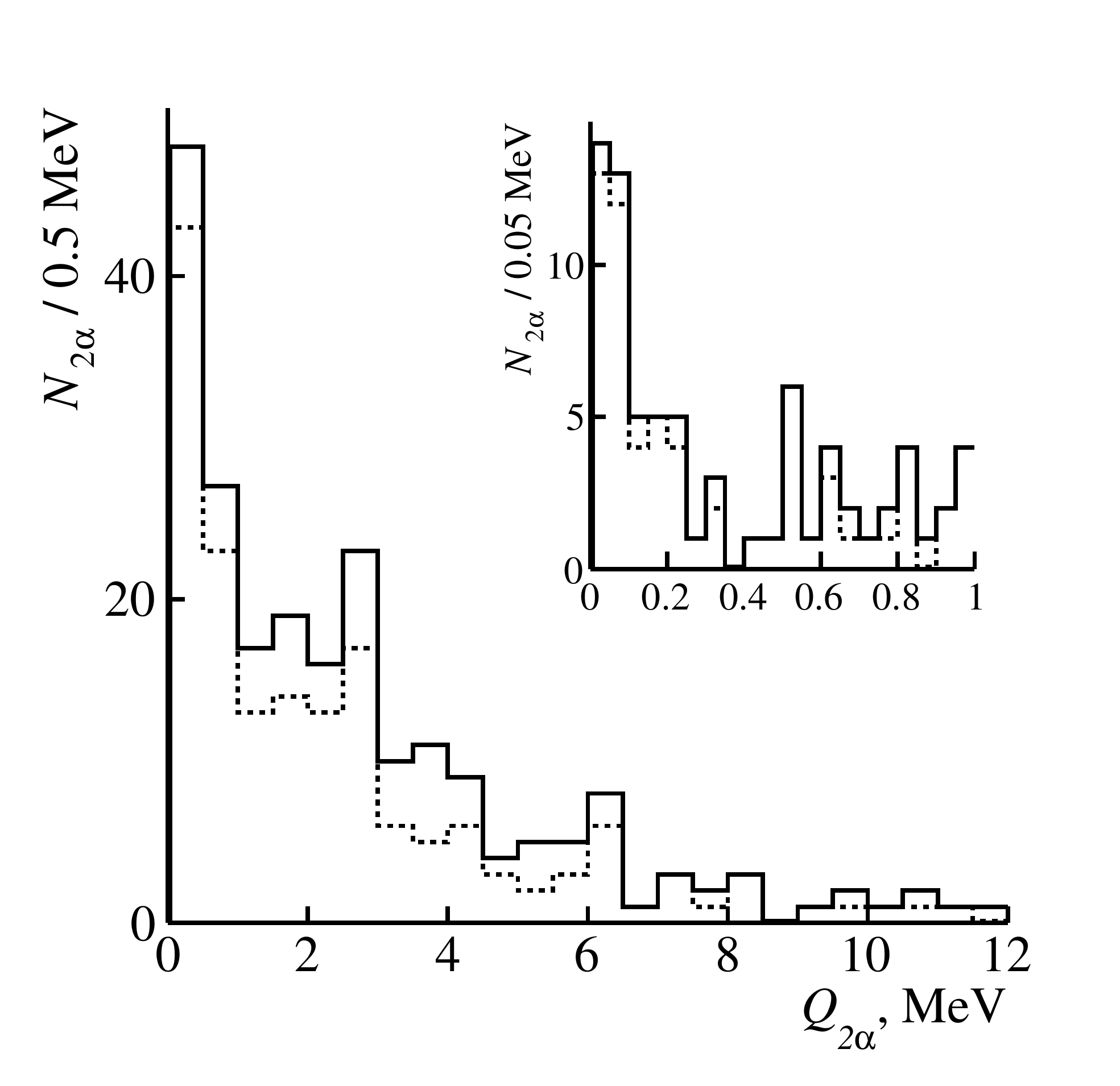}
	}
	\caption{Distribution over invariant mass $Q_{2\alpha}$ of all 2$\alpha$-pairs in events $^{14}$N $\to$ 3$\alpha$ (solid), including 60 events $^{14}$N $\to$ 3$\alpha p$ (dashed line); the inset shows enlarged region of the smallest values of $Q_{2\alpha}$.}
	\label{fig:4}       
\end{figure}

In the distribution over the invariant mass of all $2\alpha p$-triplets (Fig. 5) grouping is observed at the lowest $Q_{2\alpha p}$ values. In the region $Q_{2\alpha p}$($^{9}$B) $<$ 0.5 MeV, the average $\left\langle  Q_{2\alpha p} \right\rangle$ for 16 events is 262 $\pm$ 28 keV at RMS 120 keV which makes it possible to identify them as $^{9}$B decays. The strict condition $Q_{2\alpha}$($^{8}$Be) $<$ 0.2 MeV reduces their number to 13, which is insignificant. Thus, the contribution of $^{9}$B decays to the dissociation is $^{14}$N $\to$ 3$\alpha p$ 22 $\pm$ 9\%, and to the $^{8}$Be decays 53 $\pm$ 16\%. This contribution to the dissociation of $^{10}$B $\to$ 2$\alpha p$ nuclei is 8 $\pm$ 2\% (26/315) and 45 $\pm$ 11\% (26/58), respectively \cite{2}.

Noteworthy is the fact that the $^{9}$B contribution noticeably increases upon passing from 2$\alpha p$ to 3$\alpha p$ ensembles with a corresponding increase in the 2$\alpha p$ combinations while maintaining the ratio of the numbers $^{8}$Be and $^{9}$B. A similar behavior was noted in the study of (HS) \cite{2}. An increase in 3$\alpha$-combinations in $^{16}$O $\to$ 4$\alpha$ leads to a noticeable increase in the contribution of HS decays, and the ratio of the yields of $^{8}$Be and HS is approximately constant.

\begin{figure}
	\resizebox{0.6\textwidth}{!}{%
		\includegraphics{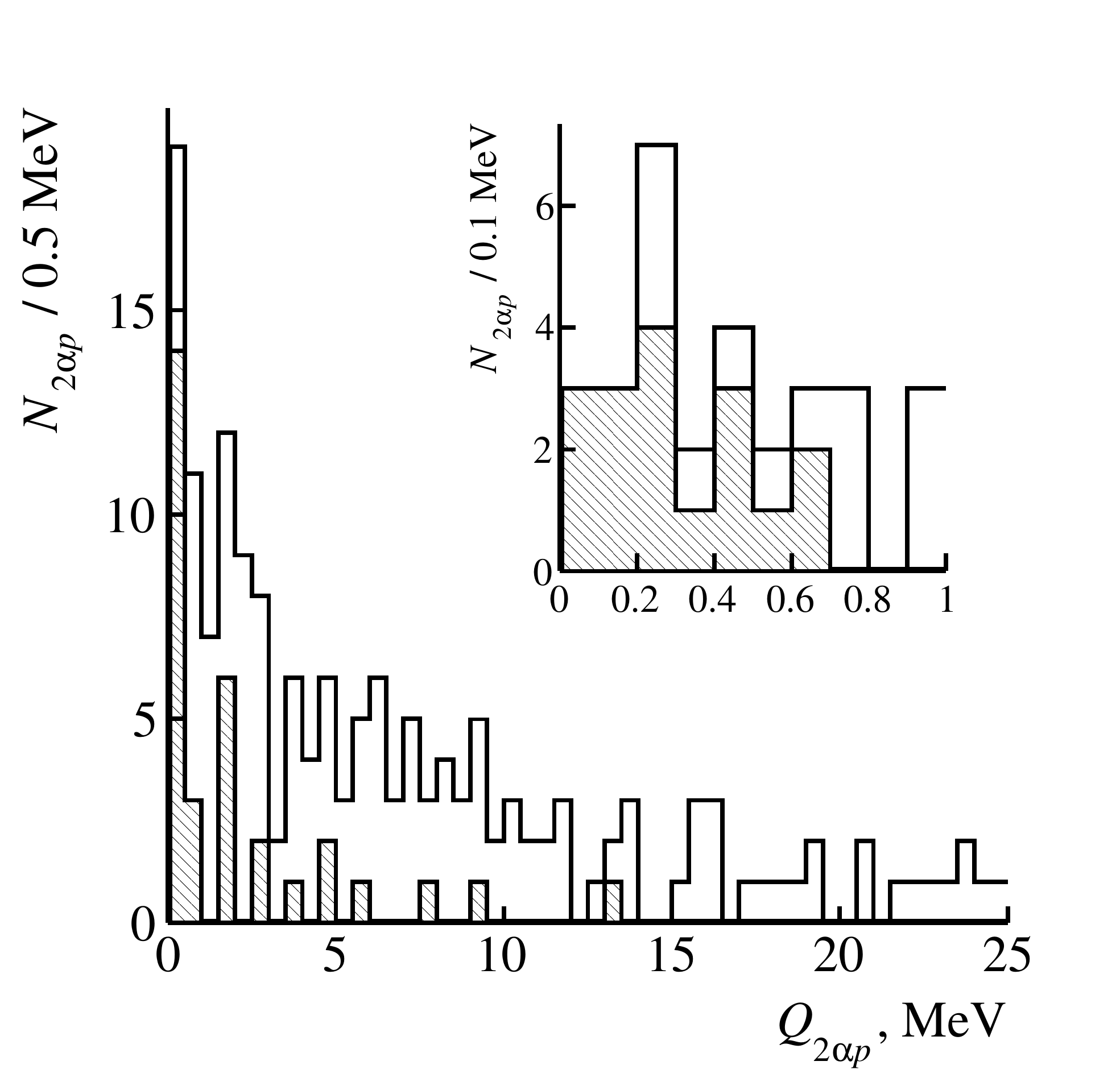}
	}
	\caption{Distribution over invariant mass $Q_{2\alpha p}$ of all 2$\alpha p$-triplets in events $^{14}$N $\to$ 3HeH (solid), including $^{14}$N $\to$ $\alpha ^{8}$Be$p$ (dashed line); inset: enlarged region of smallest values of $Q_{2\alpha p}$.}
	\label{fig:5}       
\end{figure}

\begin{figure}
	\resizebox{0.7\textwidth}{!}{%
		\includegraphics{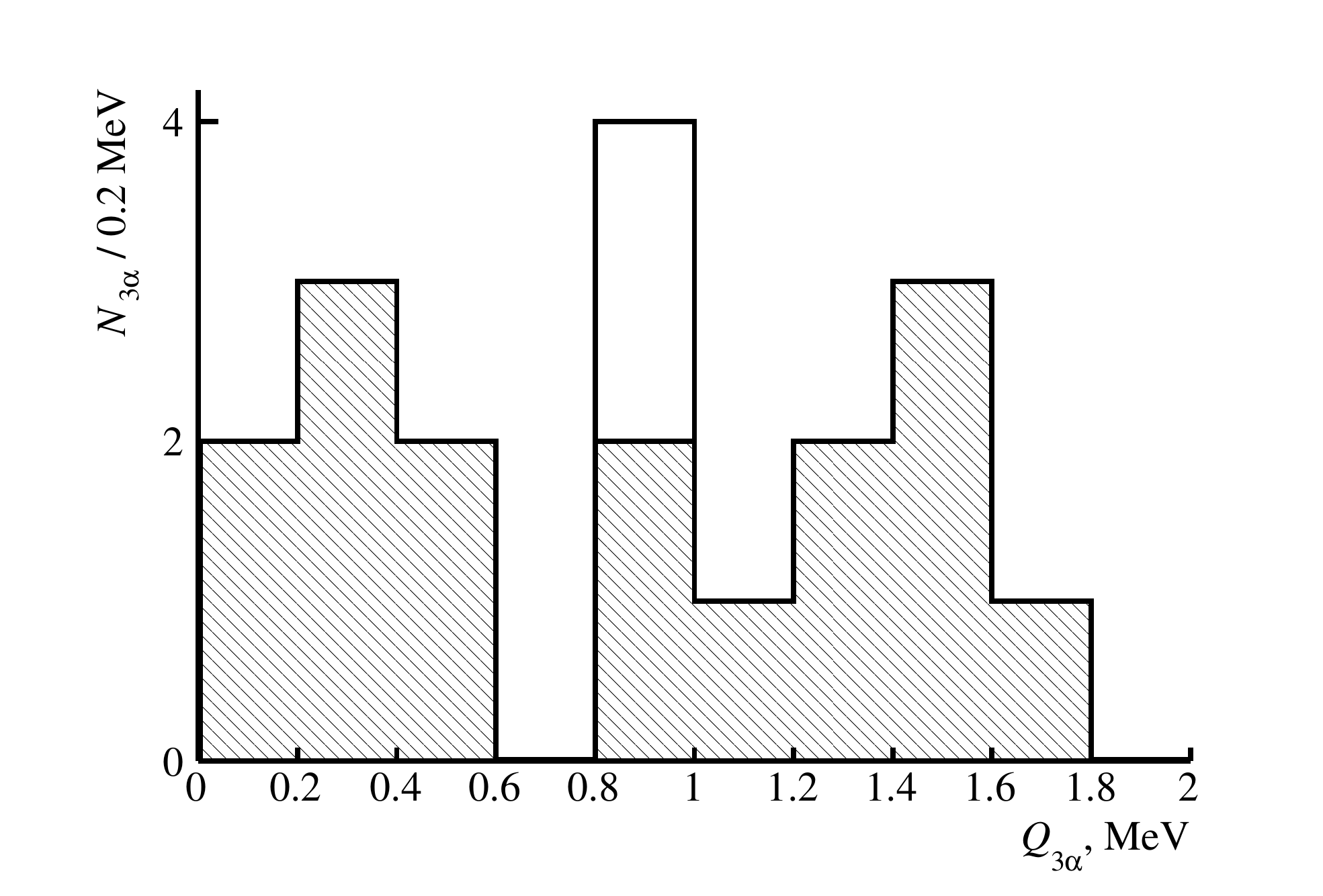}
	}
	\caption{Distribution in region of smallest values of invariant mass $Q_{3\alpha}$ for all 3$\alpha$-triplets in events $^{14}$N $\to$ 3He (solid), including $^{14}$N $\to$ $\alpha ^{8}$Be (dashed line).}
	\label{fig:6}       
\end{figure}

In the invariant mass distribution of all 3$\alpha$-triples (Fig. 6), the condition $Q_{3\alpha}$(HS) $<$ 0.7 MeV is satisfied by 7 events with $\left\langle  Q_{3\alpha} \right\rangle $ = 328 $\pm$ 60 keV at RMS 158 keV. Of these, 2 events of 3HeH can also be attributed to $^{9}$B decays. The contribution of HS decays to $^{8}$Be decays is estimated at 23 $\pm$ 10\%. Without allowing discussion of the resonance formation, the number of small 3$\alpha p$-quadruples with a small invariant mass $Q_{3\alpha p}$ motivates building up statistics (Fig. 7).

\begin{figure}
	\resizebox{0.7\textwidth}{!}{%
		\includegraphics{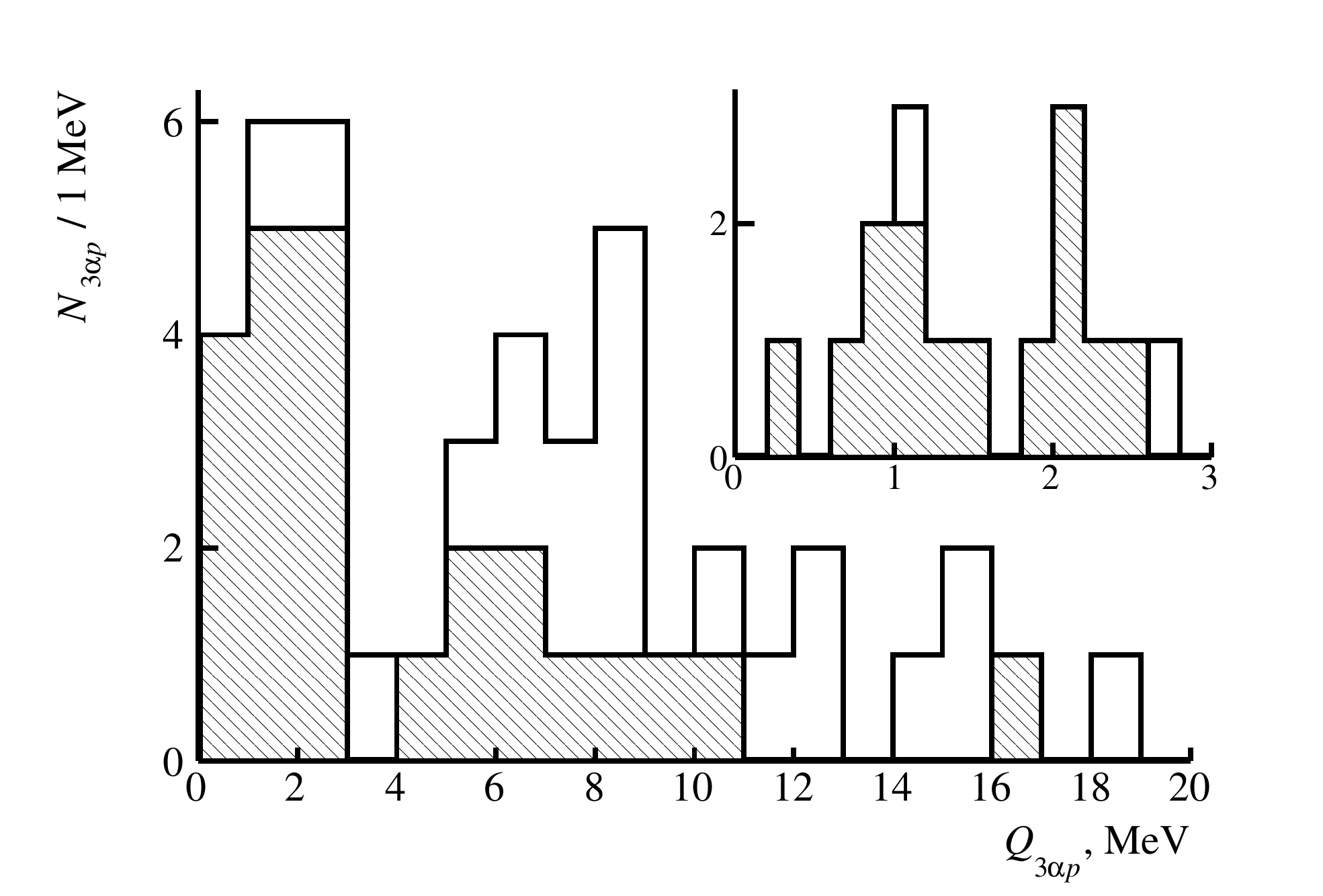}
	}
	\caption{Distribution over invariant mass $Q_{3\alpha p}$ of all 3$\alpha p$-quartets in events $^{14}$N $\to$ 3HeH (solid), including $^{14}$N $\to$ $\alpha ^{8}$Be$p$ (dashed line); inset: enlarged region of smallest values of $Q_{3\alpha p}$.}
	\label{fig:7}       
\end{figure}

\section{Conclusions}
\label{sec:3}

The analysis of dissociation of the nucleus $^{14}$N in respect to contribution of the unstable $^{8}$Be and $^{9}$B 
nuclei is presented. Primary data on the topology of the leading channels for both $^{10}$B and $^{14}$N nuclei point out 
to reflection of the $^{10}$B structural features to the leading channel $^{14}$N $\to$ 3He (+ H). The contribution of 
$^{8}$Be and $^{9}$B to the $^{14}$N dissociation has been determined. To more than half of the $^{8}$Be decays 
correspond to $^{9}$B decays. This fact also corresponds to the presence of the $^{10}$B component. A new feature is the 
identification of HS decays in the dissociation $^{14}$N $\to$ 3He (+H), which expands the idea of the universality of 
HS, previously established for the dissociation of $^{12}$C and $^{16}$O nuclei. HS decays correspond to about another 
quarter of $^{8}$Be decays not associated with $^{9}$B decays.
 
The similarity of the topology of dissociation of the nuclei $^{10}$B and $^{14}$N in the region of limiting fragmentation is based on the similarity of their structural features. Approximately equal ratios of the number of events with the formation of target fragments $^{14}$N($^{10}$B) $\to$ 3(2)He are universally determined by knocking out external neutrons and protons. The spatial distribution of the external neutrons for both nuclei appears to be twice as wide as for the external protons in both nuclei.
 
Noteworthy is the fact that no more than half of $^{8}$Be decays correspond to $^{9}$B decays, and the ratio of their number approximately coincides in the leading $^{14}$N($^{10}$B) $\to$ 3(2)HeH channels. In the case of coherent dissociation of the $^{10}$C $\to$ 2$\alpha$2$p$ nucleus, the decays of $^{8}$Be and $^{9}$B coincided completely, and their contribution was 30\% \cite{1}. Earlier, an unexpectedly large ratio of probability of coherent dissociation of $^{10}$B in the mirror channels $^{9}$B$n$ and $^{9}$Be$p$ was found amounting to 6 $\pm$ 1 \cite{6}. It seems that the $^{9}$Be nucleus itself is mainly absent as an ingredient of $^{10}$B and $^{14}$N, and instead of it there is a rarefied nuclear-molecular structure of $^{8}$Be$n$. At the same time, the $^{9}$B ground state presents such a structure initially. Being bound, the $^{8}$Be nucleus plays the role of a base in both cases. By assigning half of the $^{8}$Be decays to the $^{8}$Be$n$ structure one could restore the mirror symmetry and explain the leading of the 3(2)He channels. Since relativistic neutron experiments are difficult, a hypothesis worthy of testing in the low-energy region.

To conclude, a 3-fold increase in statistics is necessary and possible measured events $^{14}$N $\to$ 3He (+H), especially with regard to the search for resonance states 3$\alpha p$.

\end{document}